\documentstyle[aps,prl,multicol,epsf]{revtex}

\title{\bf Influence of Hydrodynamic Interactions on the Kinetics of
Colloidal Particle's adsorption}

\author{P. Wojtaszczyk$^{\dagger}$ and J.B. Avalos$^{*}$ \\
$^{\dagger}$ Departament de
F\'{\i}sica Fonamental, 
Facultat de F\'{\i}sica,\\Universitat de Barcelona,
\\Avda. Diagonal 647, E-08028 Barcelona (Spain) \\
$^{*}$ Departament d'Enginyeria
Qu\'{\i}mica, ETSEQ \\
Universitat Rovira i Virgili\\ Carretera de Salou s/n, E-43006
Tarragona (Spain)}

\begin{document}
\maketitle\parskip 2ex
\renewcommand{\theequation}{\arabic{equation}}

\begin{abstract}
The kinetics of irreversible adsorption of spherical particles onto a flat
surface is theoretically studied. Previous models, in which hydrodynamic
interactions were disregarded, predicted a power-law behavior $t^{-2/3}$
for the time dependence of the coverage of the surface near saturation.
Experiments, however, are in agreement with a power-law behavior
of the form $t^{-1/2}$. We outline that, when hydrodynamic interactions
are considered, the assymptotic behavior is found to be compatible with
the experimental results in a wide region near saturation.
\end{abstract}

PACS{68.10.Jy,02.50.+s,82.65.-i}

\begin{multicols}{2}
\setcounter{equation}{0}        

        The adsorption of colloidal particles or macromolecules such as
proteins onto adsorbing surfaces is a very common phenomenon in many
fields of Biology, Chemistry and Physics. Deposition of bacteria on teeth
or the adsorption of antibodies on living cells are examples of such
phenomena which are of great interest in medical sciences. In many cases,
the adsorption is irreversible under conditions of practical interest.  A
model system to study the adsorption process is a suspension of latex
spheres put in contact with a suitable adsorbing flat surface.  Much work
has been done for this system
\cite{lang,ada1,ada2,ada3,stprl,sprl,tar,woj1,woj2,sflu,sfl2,senf,mpar},
both on the structural properties of the adsorbed
layer and on the kinetics of the process. 
For irreversible adsorption of sufficiently light spherical particles onto
a flat 
surface, Schaaf {\em et al.}\cite{sprl} predicted a power-law behavior
$t^{-2/3}$ for the time-dependence of the coverage of the surface near
saturation.  This behavior is due to the interplay of the diffusion of
particles from the bulk and the {\em blocking effect} caused by the
saturation of the surface due to the previously adsorbed particles. This
result agrees with Brownian dynamics simulation of spheres diffusing from
a bulk solution with a constant diffusion coefficient\cite{sen}.
Experiments on the adsorption kinetics of
small spherical particles (proteins or small latex particles), however, show
a power law behavior of the form 
$t^{-1/2}$\cite{kinexp,lenov} when gravitational effects on the
particles can be ignored in front of its pure diffusion. Such a behavior
is predicted by kinetic models 
based on the Random Sequential Adsorption (RSA) filling rules\cite{stprl}.
RSA, however, ignores the physical mechanisms driving the particles to the
surface.  Thus, it seems that an important physical ingredient has been
missed in previous approaches\cite{sprl}. In this Letter, we will show
that hydrodynamic interactions between the particles and the surface
substantially modify the predicted asymptotic behavior of the system near
saturation. A simple theoretical
model taking into account the blocking effect, the diffusion of the
particles from the bulk and the hydrodynamic interactions between the
particles and the surface, allows us to predict a complex asymptotic behavior
for the time-dependence of the surface coverage near saturation. Indeed,
near saturation,
the model predicts a wide first time-domain where the time-dependence of the
coverage is dominated by the hydrodynamic interactions between the free
particles and the adsorbing surface. Remarkably, the time-dependence in
this region is compatible with the experimental findings of refs.
\cite{kinexp,lenov}. Furthermore, our model also allows to derive a second
time-domain in which the dynamics is dominated by the
blocking effect. In this terminal regime, the asymptotic time-dependence
of the coverage
is in agreement with the preditcions of Schaaf et al.\cite{sprl}.
Nevertheless, we will see that this terminal regime should not be
observed for short-range adsorption potentials, whose interaction range is
much smaller than the size of the colloidal particles.

         Despite their apparent
simplicity, the deposition processes are determined by the interplay of
various phenomena: the Brownian motion of the free particles, the
gravitational force, the dynamic interactions mediated by the solvent
(hydrodynamic interactions) and all other kinds of
interactions between free particles and the adsorbed ones, as well as
between the free particles and the wall. Irreversible adsorption
leads to non-equilibrium configurations, thus, it
cannot be studied in the framework of equilibrium statistical mecanics.
Most of the previous models\cite{lang,stprl,rsa,bal} have neglected the
effect of the solvent. They consider the particles as 
moving in dry water\cite{fey} and have focused primarily on the geometric
aspects, related to the excluded surface effects. Recently, however, the
determinant role played by the hydrodynamic behavior of the solvent was
pointed out \cite{bafa,igna1}. For instance, the theoretically predicted 
pair distribution function of the adsorbed layer 
shows significant 
deviations from the experimental curves for this function when
hydrodynamic interactions are 
ignored\cite{woj1,woj2,igna1,igna2,PhD}.  The effect of the hydrodynamic
interaction is to increase the frictional force experienced by a particle
when it approaches a flat surface. Despite its clear implication in the
kinetics of the adsorption process, the effect of the hydrodynamic
interaction in the time- dependence of the coverage near saturation has
never been analyzed before.  In the analysis proposed here we assume that
the free particles diffusing from the bulk have to cross an
{\em entropic} barrier, due to the presence of the previously adsorbed
(bound) particles in the layer, before they get trapped by an adsorbing
short range potential between the particles and the plane. The density
$\rho$ of free particles in the 
region near the wall is assumed to satisfy a diffusion equation of the
form\cite{nous}
\begin{equation}
\frac{\partial}{\partial t} \rho= -\frac{\partial}{\partial \gamma}
J(\gamma,t) =
\frac{\partial}{\partial
\gamma} \frac{D(\gamma,\theta)}{R^2} \left[\frac{\partial}{\partial \gamma}
\rho-\rho \frac{\partial}{\partial \gamma} \ln\Phi(\gamma,\theta) \right]
                \label{2} 
\end{equation}
where $\gamma \equiv z/2R$ is the dimensionless coordinate in
the $z$-direction orthogonal to the wall, $R$ being the radius of the
particles. The origin of coordinates is taken at the center of an adsorbed 
particle. In this equation,
$J(\gamma,t)$ is the flux of free particles in the vicinity of the
wall. Thus, it is in the region $0 \leq \gamma \leq 1$ that the effect of
the excluded surface due to the presence of adsorbed particles takes place. 
$\Phi(\gamma,\theta)$ is the available area for a free particle to move in at
a height $\gamma$ and at a coverage
$\theta$ of the surface\cite{wid}. $\theta$ is defined as $\theta \equiv
\pi R^2 \rho_s$, where the number of adsorbed particles per unit of area is
denoted by $\rho_s$.  The diffusion coefficient $D(\gamma,\theta)$ is
related to the mobility of the free particles. Far from the wall, the
diffusion coefficient of spheres in a dilute solution is given
simply by the Stokes-Einstein formula $D = kT/6\pi \eta R$, $\eta$ being
the viscosity of the solvent and is constant. Near the wall, however,
hydrodynamic interactions modify this behavior. Lubrication
theory\cite{Rus} shows that the mobility in the direction
orthogonal to the wall vanishes {\em linearly} with the distance between the
hydrodynamic surfaces. Thus, the diffusion coefficient behaves as
\begin{equation}
D(\gamma,\theta) = (\gamma+\delta) D_0  \label{dif}
\end{equation}
\noindent as $\gamma \rightarrow 0$. $D_0$ is a constant and $\delta \equiv
d/2R$, where $d$ stands for the repulsive range of the adsorption
potential (fig. 1). Note that the {\em hydrodynamic} wall is then shifted with
respect to the {\em adsorbing} wall due to the finite range of the adsorption
potential considered here. $\delta$ is finite but can be made arbitrarily
small 
in our model, in order to compare with previous analysis\cite{sprl}.  If
$\delta \rightarrow 0$ no particles can be adsorbed in a 
finite time due to the strength of the lubrication forces. The
hydrodynamic interaction between the free particles and the adsorbed
ones is subdominant for the motion in the direction orthogonal to the
wall\cite{Rus}, due to the fact that their surfaces move parallel to each
other when $\gamma \rightarrow 0$. As a consequence, the diffusion
coefficient is independent of the coverage $\theta$ in this limit.

        The diffusion equation (\ref{2}) contains all the relevant
phenomena driving the kinetics of particle adsorption. In the expression
between brackets, the first term stands for pure diffusion of the free
particles while the second accounts for the fact that the available area at
a given height $\gamma$ and at a given coverage $\theta$ is limited by the
presence of the adsorbed particles. Therefore, if the available area is
reduced as $\gamma$ decreases, this term acts as an effective entropic
potential tending to decrease the flux of free particles. Our model can 
be applied to a large class of systems
provided that they meet the following requirements; (i) the particles are
irreversibly adsorbed on the surface and stop moving once trapped; (ii)
diffusion dominates over gravitational effects in the dynamics of the
free particles. 

Notice that, the use of equation (\ref{2}) implies that we
describe the transport of the free  
particles across the layer of adsorbed ones without explicit calculation of
the structural properties of the layer. Brownian motion
permits the particles to explore large regions in space before they get
adsorbed. Thus, one expects that the overall adsorption process is not
determined by the local inhomogeneities of the layer of adsorbed particles
but by its global properties, in the spirit of a mean field approach \cite{nous}.

        The saturation coverage
$\theta_{\infty}$, or {\em jamming limit}, is reached when on the
adsorbing surface $(\gamma=0)$, the available surface function becomes
equal to zero. For $\theta=\theta_{\infty}$, the entropic barrier becomes
infinite, thus the adsorption of new incoming particles is
imposible. For spherical particles, the entropic potential near saturation
can be written as
\begin{equation}
\ln \Phi(\theta,\gamma) \simeq
\ln \left(\theta_{\infty}-\theta (1-\gamma^2) \right)^3      \label{pot}
\end{equation}
\noindent The form $(\theta_{\infty}-\theta)^3$ is the behavior of the
available area near saturation for irreversible adsorption of particles,
and has been first derived by Pomeau\cite{Pom}. In addition, we have
explicitly indicated the fact that, 
at a given height, the area excluded by the adsorbed spherical particles is
reduced by a factor $(1-\gamma^2)$ for spheres\cite{nous}.

        In order to describe the kinetics of the adsorption process, we
have to find the incoming flux of free particles arriving at the
adsorbing surface $J_s(t) = J(\gamma=0,t)$. Since we are interested only
in the kinetics near saturation, where the adsorption process is very slow
due to the blocking effect, we can solve eq. (\ref{2}) neglecting the
explicit time-dependence of $\rho$. We then assume that the variations in the
density profile and, thus, in the flux, adiabatically follow the changes in
the coverage through $\Phi(\theta,\gamma)$\cite{nous}. Therefore, we set
$\partial\rho/\partial t \simeq 0$ in the left hand side of eq. (\ref{2}),
implying that   
$J(\gamma,t)$ is independent of $\gamma$ and equal in fact to $J_s(t)$.
We consider here that the 
density of particles in the bulk $\rho_B$ is the control parameter and thus express
$J_s$ in terms of $\rho_B$, with the boundary conditions
\begin{eqnarray}
\rho(\gamma = 1) &=& \rho_B, \label{3}\\
\rho(\gamma = 0) &=& 0 \label{4}
\end{eqnarray}
\noindent The first boundary condition assumes that the density of bulk
particles in the vicinity of the adsorbed layer is
approximately constant due to the slow adsorption process occuring near
saturation. We thus consider a particle's reservoir
located at $\gamma = 1$ with a density  $\rho=\rho_B$ constant. The
second boundary condition stands for an irreversible adsorption: free
particles 
reaching the wall become irreversibly adsorbed and then the density of
free particles is zero at $\gamma = 0$. 
We can thus
obtain the flux of particles reaching the surface in terms of $\rho_B$ by
solving the differential equation
\begin{equation} 
J_s=-\frac{D(\gamma)}{R^2} \left[\frac{\partial}{\partial \gamma}
\rho-\rho \frac{\partial}{\partial \gamma} \ln \Phi \right] \label{5}
\end{equation}
\noindent with boundary conditions specified in eqs. (\ref{3})
and (\ref{4}). We find the following kinetic equation
\begin{equation}
\frac{\partial\rho_s}{\partial t} = -J_s = -\frac{D_0}{R^2} \rho_B I(\theta) \label{jsu2}
\end{equation}
\noindent where
\begin{equation}
I(\theta)=\frac {1}{\int_{0}^{1}\frac{D_0}{D(\gamma) \,
\Phi(\theta,\gamma)}d\gamma} \sim \frac
{1}{\int_{0}^{1}\frac{D_0}{D(\gamma)(\theta_{\infty}-\theta
(1-\gamma^2))^3}d\gamma}\; 
\label{hjsu}
\end{equation}
as $\theta \rightarrow \theta_{\infty}$. A closed equation for the
time-dependece 
of the coverage then follows by multiplying both sides by $\pi R^2$, yielding 
a generalized Langmuir equation\cite{nous}
\begin{equation}
\frac{\partial\theta}{\partial t} = K_a \rho_B I(\theta) \label{lang}
\end{equation}
\noindent where we have defined the kinetic coefficient $K_a = D_0 \pi$.
A crucial point in our analysis is that the leading contribution 
to $I(\theta)$ near 
saturation depends on the behavior of the integrand for small $\gamma$, which
allows us to use the expression of $D(\gamma)$ given in eq. (\ref{dif}).
Inserting  
this dependence in the
right hand side of eq. (\ref{hjsu}) we obtain that the adsorption rate
near saturation is proportional to 
\begin{equation}
I(\theta) \sim \frac
{1}{\int_{0}^{1}\frac{1}{(\gamma+\delta)(\theta_{\infty}-\theta
(1-\gamma^2))^3}d\gamma}        \label{eq}
\end{equation}
The asymptotics of $I(\theta)$ as given by this expression strongly
depends on the relative magnitude of $\delta$ and $\Delta \theta \equiv
(\theta_{\infty}-\theta)/\theta_{\infty}$. Clearly, when the coverage
approaches saturation, in an initial regime the condition
$(\theta_{\infty}-\theta)/\theta_{\infty} \gg \delta$ is satisfied since
$\delta$ is a constant that can be taken as arbitrarily small. In this
region, the adsorption rate is dominated by the hydrodynamic interactions
and takes the assymptotic form
\begin{equation}
I(\theta) \sim \frac{2 \Delta \theta^3}{\ln \Delta \theta/\delta^2 - 3/2}.
        \label{reg1}
\end{equation}
The range of validity of this regime is determined by the fact
that $I(\theta)$ must be positive since eq.
(\ref{lang}) describes a relaxation process in which the coverage tends
irreversibily to saturation.  Effectively, the condition $\ln \Delta
\theta/\delta^2 - 3/2 > 0$ indicates that $\Delta \theta/\delta^2 >
\exp(3/2) \sim 1$. Therefore, the crossover coverage scales as, $\Delta
\theta_c \sim \delta^2$. In this region, the time-dependence of the
coverage can 
be obtained by inserting the asymptotic behavior given in eq. (\ref{reg1})
in the right hand side of eq. (\ref{lang}). After integration we obtain
\begin{equation}
t \sim \frac{\ln \Delta \theta/\delta^2 -1}{4 \Delta \theta^2}.   \label{ech}  
\end{equation}
for $\Delta \theta/\delta^2>\exp(3/2)$. Eq. (\ref{ech}) gives an implicit
relation between the time and $\Delta \theta$. The scaling of
the crossover time $t_c$ is obtained by inserting the scaling of the
crossover coverage in this expression, giving $t_c \sim 1/\delta^4$.
Notice the fact that if $\delta \rightarrow 0$, $t_c \rightarrow \infty$,
indicating that this regime must dominate the asymptotic behavior of the
coverage near saturation. The numerator
on the right hand side of eq. (\ref{ech}) is a slowly varying function of
$\Delta \theta$. This suggests an iterative procedure to
obtain the behavior of $\Delta \theta$ for times $t \ll t_c$. Effectively,
one can write
\begin{equation}
\Delta \theta \sim \frac{1}{2 t^{1/2}} \sqrt{\ln
\left(\frac{\Delta \theta}{\delta^2} \right)} \sim \frac{1}{2 t^{1/2}} 
\sqrt{\ln \left(\frac{1}{2 \delta^2 t} \right)} \label{as}
\end{equation}
where, in deriving the last expression, a term $\sqrt{\ln \ln (\Delta \theta /
\delta^2)} \sim 1 \ll \ln (1/2\delta^2 t)$ has been neglected. Eq.
(\ref{as}) predicts a novel behavior for the time-dependence of the coverage
near saturation dominated by the hydrodynamic interactions between the
free particles and the wall. Such a behavior differs from that found from 
the RSA model\cite{Ps} as well as from that predicted by the model
incorporating the 
diffusion of the particles from the bulk\cite{sprl}.

        The terminal regime $t \gg t_c$ or, equivalently, $\Delta
\theta/\delta^2 < 1$ can also be obtained from eq.
(\ref{eq}). The adsorption rate in this case is dominated by
the blocking effect and obeys a different asymptotics of the form
\begin{equation}
I(\theta) \sim \delta \, (\theta_{\infty}-\theta)^{5/2} \label{reg2}
\end{equation}
Notice that the right hand side of this equation vanishes as $\delta
\rightarrow 0$. From eq. (\ref{lang}) and (\ref{reg2}) one arrives at the
$\Delta \theta \sim t^{-2/3}$
behavior as found by Schaaf {\em et al.}\cite{sprl}.
Therefore, the model proposed here is also able to reproduce Schaaf's
regime\cite{sprl} when the adsorption kinetics is dominated by the
blocking effect. However, due to the fact that the crossover time between
the two regimes scales as $\delta^{-4}$ and tends to infinity as $\delta
\rightarrow 0$, it suggests that this regime is never observed for short
range adsorption potentials. 

        In summary, we have explicitly discussed a model where the
hydrodynamic interactions are included, in addition to other physical
mechanisms like diffusion and blocking effect, which are relevant to
describe the adsorbing rate. Hydrodynamic
interactions play a crucial role in the adsorption kinetics and cannot be
avoided in any experimental work on this process. Previous models ignore
the physical mechanisms driving the particles to the surface (see
ref.\cite{stprl} and related).  Schaaf {\em et al.}\cite{sprl}, by taking
into account the diffusion of the particles from the bulk as well as the
blocking effect, made a significant step in the description of adsorption
kinetics. However, the behavior predicted by Schaaf {\em et al.} has never
been observed. Experimental results\cite{kinexp,lenov} suggest a behavior
near saturation compatible with a power law $\Delta \theta \sim
t^{-1/2}$. An important conclusion that can be drawn from the present work
is that the regime predicted by 
Shaaf {\em et al.} should not be observed for short range adsorption
potential. The most 
relevant result of this Letter is, however, to have shown that the
inclusion of 
hydrodynamic interactions leads to a behavior near saturation 
compatible with the experimental findings, in view of the slow behavior of
the logarithmic factor in eq. (\ref{as}). As it has already been pointed
out for structural aspects of colloidal particles' adsorption onto
solid surfaces\cite{bafa,igna1}, the results of the present work
stress the importance of the hydrodynamic interactions also in the
kinetics of the process.

P. Wojtaszczyk would like to acknowledge Professor J.M. Rubi, G. Gomila
and J.M.G. Vilar for fruitful discussions and suggestions.  This work has
been supported by the "Ministerio de Educacion y Sciencia" of Spain
(contract No. SB95-B41700129).

\section*{Figure captions}

Schematic representation of the adsorption process. The adsorbing (free)
particles diffuse in the bulk and get finally adsorbed at a distance $d$
(arbitrarily small) from the wall. The origin of the dimensionless
coordinate $\gamma$ is taken in the plane defined by the centers of the
adsorbed particles.

\end{multicols}

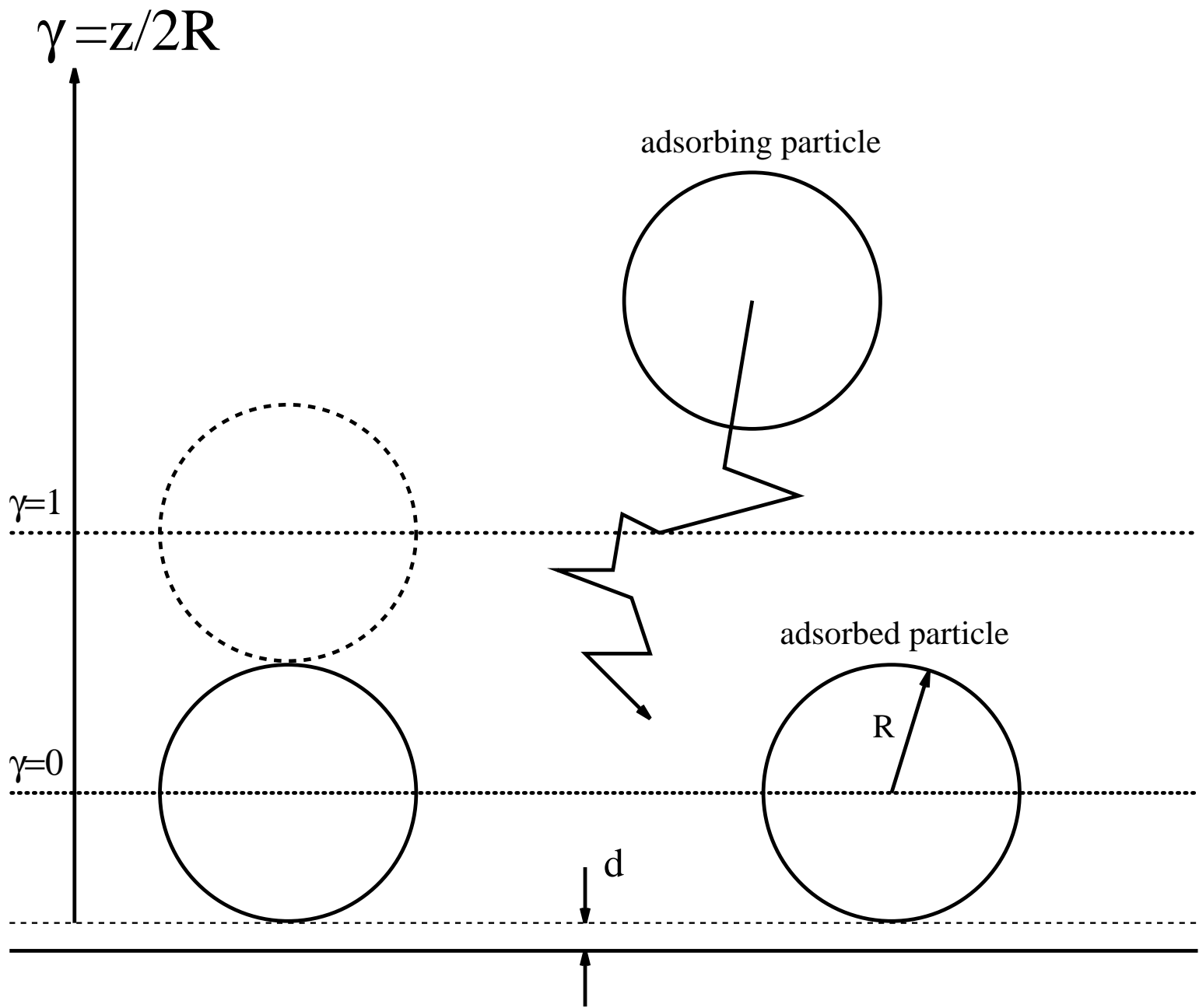
\end{document}